\documentclass[aps,twocolumn,,superscriptaddress,showpacs,showkeys,amsmath,amssymb,floatfix]{revtex4}
\usepackage{graphicx}
\usepackage{epsfig}
\usepackage{dcolumn}
\usepackage{bm}
\usepackage{amssymb}
\usepackage{dsfont}
\usepackage{amsmath}
\newcommand{\be}{\begin{equation}}
\newcommand{\ee}{\end{equation}}
\newcommand{\bea}{\begin{eqnarray}}
\newcommand{\eea}{\end{eqnarray}}

\newcommand{\hatm}{\hat m}
\newcommand{\tC}{{\tilde C}}

\newcommand{\pro}{\partial}
\newcommand{\der}{\partial}

\newcommand{\ba}{\begin{array}}
\newcommand{\ea}{\end{array}}

\newcommand{\nn}{\nonumber}

\newcommand{\uast}{u^*}

\begin{document}

\title{Knot topology in QCD}
\bigskip
 \author{L. P. Zou}
\affiliation{Institute of Modern Physics, Chinese Academy of Sciences,
 Lanzhou 730000, China}
 \affiliation{
School of Nuclear Science and Technology, Lanzhou University, Lanzhou 730000, China}
\affiliation{
University of Chinese Academy of Sciences, Beijing 100049, China}
 \author{P. M. Zhang}
\affiliation{Institute of Modern Physics, Chinese Academy of Sciences,
 Lanzhou 730000, China}
  \affiliation{Research Center for Hadron and CSR Physics,
 Lanzhou University and Institute of Modern Physics of CAS, Lanzhou 730000, China}
\author{D. G. Pak}
\affiliation{Institute of Modern Physics, Chinese Academy of Sciences,
 Lanzhou 730000, China}
\affiliation{Lab. of Few Nucleon Systems,
 Institute for Nuclear Physics, Ulugbek, 100214, Uzbekistan}

\begin{abstract}
We consider topological structure of classical vacuum solutions in
quantum chromodynamics.
Topologically non-equivalent vacuum configurations are
classified by non-trivial second and third
homotopy groups for coset of the color group $SU(N)~(N=2,3)$
under the action of maximal Abelian stability group.
Starting with explicit vacuum knot configurations we study
possible exact classical solutions.
Exact analytic non-static knot solution in a simple $CP^1$ model
in Euclidean space-time has been obtained. We construct an ansatz
based on knot and monopole topological vacuum structure for searching
new solutions in $SU(2)$ and $SU(3)$ QCD.
We show that singular knot-like solutions in QCD in Minkowski space-time
can be naturally obtained from knot solitons in integrable
$CP^1$ models. A family of Skyrme type low energy effective theories
of QCD admitting exact analytic solutions with non-vanishing
Hopf charge is proposed.
\end{abstract}
\vspace{0.3cm}
\maketitle

\vspace{2mm}

\section{Introduction}

Topological structure of classical solutions in $SU(N)$ Yang-Mills
theory implies numerous physical manifestations
in such important phenomena in quantum chromodynamics (QCD)
as the chiral symmetry breaking and confinement \cite{shuryak98}.
The most attractive mechanism of the confinement is based on
the Meissner effect in dual color superconductor
\cite{nambu, mandelstam, polyakov77} where monopole
vacuum condensate is generated dynamically due to
quantum corrections \cite{savv,NO78,prd02}.
Dyons represent alternative topological
defects which may play an important role as well as monopoles
in description of the confinement at zero and finite temperature
\cite{shuryak13}.
Whereas the instanton and monopole solutions
correspond to non-trivial topological Chern-Simons and monopole charges,
the topological knot configurations with a non-zero Hopf charge
represent another topological objects which become
essential in various applications in standard QCD and
effective Skyrme type theories of QCD in low energy region
\cite{skyrme,skyrmereview}.
It has been found that knot solitons could be good
candidates for description of glueball states which can be treated
as excitations over the condensed vacuum \cite{FNnature,choprl01}.

The rich topological structure of QCD as a gauge theory
is conditioned by the presence of non-trivial
homotopy groups $\pi_k(SU(N)/H)$, where the stability subgroup $H$
determines possible coset spaces with different topological properties.
The homotopy group $\pi_3(SU(N))=Z$ describes topological classes
of instanton field configurations corresponding to
topological Pontryagin index \cite{bpst,callan,jackiw}.
It is well known that instantons realize tunneling between topologically non-equivalent
vacuums and provide dominant contribution to chiral symmetry breaking.
Another example of manifestation of non-trivial topology in QCD is provided by
the second homotopy group $\pi_2(SU(3)/U(1)\times U(1))=Z\times Z$
which implies Weyl symmetric structure of vacuum and singular
monopole solutions in $SU(3)$ QCD \cite{goddard,choprl80,choplb82,weylsym12}.

A nice feature of quantum chromodynamics is that gauge connection (potential)
allows natural implementation of the color vector
$\hat n$ in adjoint representation of $SU(N)$ within the formalism of gauge invariant
Abelian decomposition suggested first in \cite{choprl80, choprl81, duan,zhang}
and developed further in \cite{FNdec,fadd1,shab}.
The color vector $\hat n$ corresponds to generators of the Cartan subalgebra of
Lie algebra $su(N)$ and gives a suitable tool for description
of whole topological structure of the gauge theory.
A crucial observation has been made that the
classical vacuum in QCD can be explicitly constructed in terms
of the color vector $\hat n$ \cite{baal,choplb07}.
{\it This immediately implies that classical vacuum is strongly degenerated
and all topologically non-equivalent vacuums are classified by
non-trivial homotopy groups $\pi_{2,3}(SU(N)/H)$}.
In particular, in the case of $SU(3)$ QCD
it has been shown that classical vacuum possesses a
non-trivial Weyl symmetric structure
described by the second homotopy group $\pi_2(SU(3)/U(1)\times U(1))$ \cite{weylsym12}.
It should be stressed, that the color vector $\hat n$ represents
pure topological degrees of freedom, so that we have still a standard QCD.
In low energy region, in effective QCD theories like a generalized
Faddeev-Skyrme model \cite{FNnature}, the vector $\hat n$ becomes
dynamical. The knowledge of the classical vacuum structure
allows to study vacuum excitations in search of possible
finite energy topological solutions and make further steps towards
understanding fundamental properties of QCD at quantum level.

In the present paper we consider first the
topological structure of the classical vacuum
in $SU(N),~ (N=2,3)$ QCD and study possible manifestations of topological
properties related with the homotopy groups
$\pi_{2,3}(SU(3)/U(1)\times U(1))$ and $\pi_{2,3}(SU(2)/U(1))$.
Starting with known exact knot solutions from the integrable sector
of $CP^1$ models \cite{nicole, AFZ} we will construct a vacuum with knot
topology and obtain new analytic classical solutions
in $CP^1$ model, standard QCD (in Euclidean and Minkowski space-time)
and effective Skyrme type theory.
The paper is organized as follows.
In Section II we describe the general topological structure
of the classical vacuum in $SU(2)$ and $SU(3)$ QCD.
In Section III we consider a simple $CP^1$ model which
can be treated as a restricted QCD
with one field variable $\hat n$.
Exact non-static knot like solution with a finite Euclidean
action has been found. Section IV deals with an ansatz for possible
topological solutions based on classical vacuum made of $\hat n$ with
general topology.
In Section V we present analytic singular knot like solutions
in QCD in Minkowski space-time. A family of generalized
Skyrme type effective theories admitting exact solutions
with non-trivial Hopf numbers is proposed in Section VI.

\section{Topological structure of classical vacuum in QCD}

\subsection{Topology of $SU(2)$ QCD vacuum}

It has been shown that knot configurations providing minimums
of the energy functional in Faddeev-Skyrme model may correspond to
vacuums of QCD in maximal Abelian gauge \cite{baal}.
Later it has been proved that topologically non-equivalent
classical vacuums in pure QCD can be constructed explicitly in terms
of a color vector $\hat n^a ~ (a=1,2,3)$ \cite{choplb07}. By this, the
vacuum pure gauge fields $\vec A_\mu$ with different Chern-Simons
numbers are in one-to-one correspondence
with color fields $\hat n$ of respective Hopf charges.

One should stress, that the color vector $\hat n $ in $CP^1$
models (as well as in Faddeev-Skyrme theory) represents dynamic field variable
whereas in QCD the vector field $\hat n$ contains only pure topological degrees
of freedom.
The most appropriate way how to implement the topological degrees of freedom of $\hat n$
into the gauge potential while keeping a standard QCD theory is provided by
Cho-Duan-Ge gauge invariant Abelian projection \cite{choprl80, choprl81, duan}
\bea
&& \vec A_\mu =  A_{\mu} \hat n + \vec C_\mu+ \vec X_\mu \equiv \hat A_\mu+\vec X_\mu, \label{Adec}
\eea
where $A_\mu$ and $\vec X_\mu$ are the Abelian
and off-diagonal gauge potentials,
$\hat A_\mu$ is a restricted part of the gauge potential,
and $\vec C_\mu \equiv -\dfrac{1}{g}\hat n \times \pro_\mu \hat n$ is
a magnetic potential.
For simplicity we put the coupling constant $g$ equal to one.
The vector $\hat n$ has a natural origin in the
mathematical structure of the gauge theory, it
is defined on the coset $G/H$ where the stability group
$H$ is defined by Cartan subalgebra generators of the Lie algebra $\mathfrak{g}(G)$.
Notice, that there is another type of Abelian decomposition proposed in \cite{FNdec,fadd1,shab}
which treats the color vector $\hat n$ as a part of the whole gauge potential. So that,
such a decomposition leads to a theory different from the
standard QCD already at classical level \cite{evslin}.

The magnetic field strength $\vec H_{\mu\nu}$ constructed from the magnetic
gauge potential $\vec C_\mu$ defines the scalar magnetic field $H_{\mu\nu}$
\bea
&& \vec H_{\mu \nu} = \pro_\mu \vec C_\nu - \pro_\nu \vec C_\mu
+ \vec C_\mu \times \vec C_\nu \equiv H_{\mu\nu} \hat n . \label{vecH}
\eea
The magnetic field $H_{\mu\nu}$ defines a closed differential 2-form
$H=dx^\mu \wedge dx^\nu H_{\mu\nu}$ which implies the existence of
dual magnetic potential $\tC_\mu$
\bea
&& H_{\mu\nu} = \pro_\mu \tC_{\nu }-\pro_\nu \tC_{\mu }. \label{dualpot}
\eea

An explicit construction of the classical vacuum
of QCD in terms of knot configurations of $\hat n$
has been found first in \cite{choplb07}
\bea
\vec A_\mu^{vac}=-\tC_\mu \hat n+\vec C_\mu. \label{Avac}
\eea
This relation establishes connection between a pure gauge potential
and color vector field $\hat n$ and implies that the
classical vacuum configurations can be described by topologically
non-equivalent classes of the color field $\hat n$.
Namely, the topological classes
of $\hat n$ are determined by two homotopy groups, $\pi_2(SU(2)/U(1))$ and
$\pi_3(SU(2)/U(1))=\pi_3(S^2)$.
The first one describes monopole configurations,
whereas the second homotopy describes Hopf mapping
$\hat n: S^3 \rightarrow S^2$ (we assume that the space $R^3$
is compactified to a three dimensional sphere $S^3$).
So that, all topological non-equivalent classical vacuums
are classified by Hopf, $Q_H$, and monopole, $g_m$, charges
\bea
Q_H&=& \dfrac{1}{32 \pi^2}\int d^3x \epsilon^{ijk} \tC_i H_{jk} , \nn \\
g_m&=&\int_{S^2} \vec H_{ij} \cdot \hat n \, d \sigma^{ij}.
\eea
One can show that Hopf number
equals to the Chern-Simons number for vacuum gauge field
configurations $A_\mu^{vac}$ constructed from $\hat n$.

To study possible exact solutions in QCD and QCD effective theories
we will consider explicit expressions for the color vector $\hat n$ with a given knot topology.
In particular, we will use known exact analytic knot solutions found in special
integrable models.
Let us recall first an explicit construction of a simple knot configuration of $\hat n$
as a mapping $S^3 \rightarrow S^2$ with unit Hopf charge.
Surprisingly, such a simple construction leads directly to
exact knot solutions found in $CP^1$ integrable models.
Using stereographic projection it is convenient to parameterize
the target space $S^2$ by a complex field $u\in C^1$
\bea
\hat n &=& \dfrac{1}{1+u \uast}
 \left (\ba{c}
  u+\uast\\
  -i (u-\uast)\\
 u \uast-1\\
            \ea
           \right ), \label{nstereo}
\eea
A three-dimensional sphere $S^3$ is given by embedding
into $R^4$ as follows
\bea
|z_1|^2+|z_2|^2=1
\eea
where $z_1, z_2$ are complex coordinates on the complex plane $C^2$.
The Hopf mapping with the Hopf charge $Q_H=1$ is determined by the following equation
\bea
u=\dfrac{z_1}{z_2}. \label{uzz}
\eea
Starting with a given color vector $\hat n$
one can define the magnetic field $H_{\mu\nu}$ explicitly in terms of the complex field
$u$
\bea
H_{\mu\nu}&=&\epsilon^{abc} \hat n^a \der_\mu \hat n^b \der_\nu \hat n^c= \nn \\
 && \dfrac{-2 i}{(1+|u|^2)^2} (\der_\mu u \, \der_\nu \uast-\der_\nu u \,\der_\mu \uast).
 \eea
The dual magnetic potential $\tC_\mu$, (\ref{dualpot})
is written through the complex $SU(2)$ doublet $\zeta=(z_1,z_2)$ as follows
\bea
 \tC_\mu  &=&-2 i \zeta^\dagger \der_\mu \zeta.
\eea

For physical implications an explicit realization of Hopf mapping
$\hat n$ depends on a specified model and topology of four-dimensional
space-time. The three-dimensional sphere
$S^3$ in the above definition of Hopf mapping can be related to the physical
space-time by several ways. We will consider exact solutions in
two  cases: the first case, when the sphere $S^3$ is treated as embedding into
four-dimensional Euclidean space-time $R^4$ (Section III, IV),
and the second one, when $S^3$ is obtained from the real physical space $R^3$
by suitable compactification imposing boundary conditions at space infinity
for physical fields (Section V, VI). In each case the different realizations
of the sphere $S^3$ lead to different topological field configurations.

\subsection{Topology of $SU(3)$ QCD vacuum}

The consideration of the topological vacuum structure
of $SU(2)$ QCD can be generalized to the case of $SU(3)$
gauge group which possesses a more rich topology.
The $SU(3)$ gauge invariant Abelian projection
is defined as follows \cite{choprl80, choprl81}
\bea
&& \vec A_\mu =\hat A_\mu + \vec X_\mu, \nn \\
&& \hat A_\mu = A^r_{\mu} \hatm_r + \vec C_\mu, \nn \\
&& C_\mu^a =-f^{abc} \hatm^{b}_r \pro_\mu \hatm^{c}_r \equiv
 -(\hatm_r \times \pro_\mu \hatm_r)^a,  \label{Adec2}
\eea
where two color octet fields $\hatm_r,~r=(3,8)$
correspond to Cartan subalgebra generators, i.e.,
two generators of the Abelian subgroups
$U_3(1),\, U_8(1)$.
The group manifold of $SU(3)$ can be described
within the formalism of fiber bundle
by several ways. One can realize $SU(3)$ manifold as
a fiber bundle with the base $S^5$ and a fiber $S^3$.
This is a generalization of the Hopf fibering $S^1\rightarrow S^3 \rightarrow S^2$
to the case of $SU(3)$ group.
Another possibility is to consider the group manifold $SU(3)$ as a fiber bundle
over the coset $M^6 \equiv SU(3)/U(1)\times U(1)$
with a fiber as torus $T^2=S^1 \times S^1$.
The coset space $M^6$ itself can be treated as
a bundle which is locally isomorphic
to $CP^2\times CP^1$. We will consider
the last realization of $SU(3)$ manifold because
of its importance in connection with the confinement phenomenon in QCD.

Due to presence of the non-trivial homotopy groups
\bea
&&\pi_2(SU(3)/U(1)\times U(1))=\pi_2(CP^2)\times \pi_2(CP^1)  \nn \\
 &&~~~~~~~~~~~~~~~~~=Z\times Z, \nn \\
&& \pi_3(SU(3)/U(1)\times U(1))=\pi_3(CP^1)=Z,
\eea
one concludes that nonequivalent topological classes of
color vector fields $\hat n_{3,8}$ are classified
by the Hopf number $Q_H$ and two monopole charges $(m,n)$.
As we will see later, the Hopf number is actually determined
by two integer winding numbers as well.

Let us construct vacuum configurations in terms
of independent complex field variables in
analogy with the case of $SU(2)$ QCD.
One needs to parameterize the coset $M^6$ by three complex coordinates.
To do this one should express the color vectors $\hat n_{3,8}$ in terms
of two complex triplet fields $\Psi, \Phi$ which are projective coordinates
on almost complex homogeneous manifold $M^6$.
Let us first express the lowest weight vector
$\hatm_8^a$ introducing a complex triplet
field $\Psi$ that parameterizes the coset
$  CP^2 \simeq SU(3)/SU(2)\times U(1)$ with a maximal stability
subgroup
\bea
&& \hatm_8^a = -\dfrac{\sqrt 3}{2}\bar \Psi \lambda^a \Psi, \nn \\
&& \bar \Psi \Psi=1. \label{par1}
\eea
The definition of the vector $\hat m_8$ is consistent with the
normalization condition and symmetric $d-$product operation in
the Lie algebra of $SU(3)$
\bea
&& \hatm_8^2=1, ~~~~~~~~d^{abc}\hatm_8^b\hatm_8^c=-\dfrac{1}{\sqrt 3} \hatm_8^a.
\eea
To construct a second Cartan vector $\hatm_3$ orthogonal
to $\hatm_8$ it is convenient to define projectional
operators
\bea
&& P_{\parallel}^{ab}=\hatm_8^a \hatm_8^b ,~~~~~ P_\bot^{ab} = \delta^{ab} - \hatm_8^a \hatm_8^b.
\eea
 With this the vector $\hatm_3$ can be parameterized as follows
\bea
 \hatm_3^a &=& P_\bot^{ab} \bar \Phi \lambda^b \Phi
=\bar \Phi \lambda^a \Phi +\dfrac{1}{2} \bar \Psi \lambda^a \Psi, \label{par2}
\eea
where we have introduced a second complex triplet field $\Phi$.
The definition of color vectors $\hat m_{3,8}$ by Eqs. (\ref{par1}),(\ref{par2})
is invariant under local $\tilde U(1)
\times \tilde U'(1)$ gauge transformations
\bea
&& \Psi \rightarrow \exp [i \tilde \alpha(x) ] \Psi, \nn \\
&& \Phi \rightarrow \exp [i \tilde \alpha'(x)] \Phi, \label{dualsymm}
\eea
which represent explicitly the dual Abelian magnetic symmetry in $SU(3)$ QCD.
The dual magnetic potentials $\tilde C^r_{\mu}$ can be constructed explicitly
by means of the complex fields
\bea
&& \tC^3_{\mu} = 2i (\bar \Phi \pro_\mu \Phi +\dfrac{1}{2} \bar \Psi
 \pro_\mu \Psi), \nn \\
&& \tC^8_{\mu}=2 i (-\dfrac{\sqrt 3}{2} \bar \Psi \pro_\mu \Psi). \label{dualC}
\eea
One can verify that $\hatm_r$ satisfy the following
relations
\bea
&& \hatm_r^a \hatm_s^a=\delta_{rs},~~~~~ d^{abc} \hatm_r^b \hatm_s^c = d_{rsq} \hatm_q^a , \label{ortcond}
\eea
which imply the orthogonality condition $\bar \Psi \Phi =0$
for the complex fields.

For given complex fields $\Psi=(\psi_1,\psi_2,\psi_3)$ and
$\Phi=(\phi_1,\phi_2,\phi_3)$ one can introduce complex projective coordinates
\bea
&& u_1=\dfrac{\psi_1}{\psi_2}, ~~~~~~~~~~~~ u_2=\dfrac{\psi_3}{\psi_2}, \nn \\
&& v_1=\dfrac{\phi_1}{\phi_2}, ~~~~~~~~~~~~ v_2=\dfrac{\phi_3}{\phi_2}.
\eea
With this one can obtain explicit parametrization for
the color vector $\hat m_8$
\bea
\hat m_8&=&-\dfrac{\sqrt 3}{2(1+|u_1|^2+|u_2|^2)} \cdot \nn \\
&&
 \left (
 \ba{c}
   u_1+\uast_1\\
   i (u_1-\uast_1)   \\
   |u_1|^2-1   \\
   \uast_1 u_2+\uast_2 u_1   \\
   -i(\uast_1 u_2-\uast_2 u_1 )   \\
    u_2+\uast_2   \\
   -i(u_2-\uast_2)   \\
   \dfrac{1}{\sqrt 3}(1+|u_1|^2-2 |u_2|^2)   \\
   \ea
       \right ).
\eea
An explicit expression for the vector $\hat m_3$ can be obtained
by using Eqn. (\ref{par2}) in a similar manner.
Notice, that due to the orthogonality condition
$\bar \Psi \Phi=0$ one has an additional constraint
\bea
&&1+\uast_1 v_1+\uast_2 v_2=0.
\eea
So that, we can choose three independent complex coordinate
functions, for instance, $u_1,u_2, v_1$,
which contain the whole information on the
topology of the space $SU(3)/U(1)\times U(1)$.
This can be useful in search of essentially $SU(3)$ solutions
with various combinations of Hopf and monopole charges
corresponding to topologies of the fields $u_1,u_2, v_1$.

\section{Exact non-static solution in $CP^1$ model}

The Lagrangian of a simple $CP^1$ model with $SU(2)$ triplet
field $\hat n$ coincides formally with the
Lagrangian of the restricted QCD with a
vanishing "electric" gauge potential $A_\mu$
\bea
{\cal L}=-\dfrac{1}{4} H_{\mu\nu}^2(\hat n). \label{LagrCP1}
\eea
We use four-dimensional spherical coordinate system
with Euler angles $(0\leq\theta\leq \pi,~~0\leq\phi,~\psi \leq 2\pi)$
\bea
&& x_1= \rho \cos \dfrac{\theta}{2} \cos \phi, \nn \\
&& x_2= \rho \cos \dfrac{\theta}{2} \sin \phi, \nn \\
&& x_3= \rho \sin \dfrac{\theta}{2} \cos \psi, \nn \\
&& x_4= \rho \sin \dfrac{\theta}{2} \sin \psi.  \label{sphercs}
\eea
To define Hopf fibering one introduces complex
coordinates $z_1, \, z_2$ in the complex plane $C^2$ ( $R^4$)
\bea
z_1&=&x_1+i x_2, \nn \\
z_2&=&x_3-i x_4.
\eea
A three-dimensional sphere with the radius $\rho$ is defined
by the equation
\bea
x_1^2+x_2^2+x_3^2+x_4^2=\rho^2.
\eea
The Hopf mapping with a unit Hopf charge is defined by
\bea
u=\dfrac{z_1}{z_2}=\cot \frac{\theta}{2} \, e^{i (\phi+\psi)}. \label{uzz2}
\eea
Surprisingly, that simple configuration with a Hopf number $Q_H=1$
represents an exact solution to the equation of motion for the
Lagrangian (\ref{LagrCP1}) when the field $\hat n$ is defined on
three-dimensional sphere $S^3$ with a fixed radius $\rho = const$
\bea
&& \der_\mu (\sqrt g \der_\nu u H^{\mu\nu})=0.
\eea
For higher Hopf numbers the solution is provided by the following
ansatz
\bea
u=g(\theta) e^{i (m \psi+n \phi)}.
\eea
Substituting the ansatz into Euler-Lagrange equations
corresponding to the Lagrangian ${\cal L}$ one obtains
the following solution \cite{ferr2004}
\bea
&& g(\theta)= \nn \\
&& \sqrt{-1+\dfrac{1}{c_2+\dfrac{c_1 \log [m^2+n^2-(m^2-n^2)\cos \theta]}{m^2-n^2}}}. \nn \\
&&c_1=-\dfrac{m^2-n^2}{\log [m^2]-\log [n^2]}, \nn \\
&&c_2=-\dfrac{\log [2 n^2]}{\log [m^2]-\log [n^2]}, \label{sol1}
\eea
where the integration constants $(c_1, \, c_2)$
are fixed by the boundary conditions
for the vector $\hat n$
\bea
&&\hat n(\theta=\pm \pi)=(0,0,\pm 1).
\eea
The corresponding Hopf charge is determined by two winding numbers
($m,~n$)
\bea
Q_H=mn.
\eea
In the case $m=n=1$ the solution reduces to the special one, Eq. (\ref{uzz2}).
Notice, the solution (\ref{sol1})
can be treated as a "static" soliton in a sense that it
does not depend on the radius $\rho$.
This solution coincides exactly with the solution obtained
in \cite{ferr2004} with the field $\hat n$ defined on three-dimensional
sphere of finite radius $r_0$. In other words, the topology of the
Euclidean space-time is chosen to be $S^3 \times R^1$.
The presence of the dimensional parameter
$r_0$ allows to overcome the restrictions of the Derrick theorem to existence
of finite energy stable solutions.
Notice, that solution (\ref{sol1}) does not
admit analytical continuation to the Minkowski space-time.
Approximate solutions for similar knotted soliton on three-dimensional sphere
in Fadeev-Skyrme model had been constructed in \cite{ward}.

In our approach we do not restrict the Euclidean four-dimensional
space-time to the topology $S^3 \times R^1$.
To find a non-static solution to equations of motion of QCD in
Euclidean space-time $R^4_E$ we apply the same ansatz
which is used in \cite{ferr2004}
\bea
u=f(\rho,\theta)\exp[i(m \phi +n \psi)].
\eea
After proper changing variable $g(\rho,\theta)=\dfrac{1}{1+f^2(\rho,\theta)}$
one results in a linear partial differential equation
\bea
&&\rho\der_\rho(\rho \der_\rho g)+\dfrac{4}{A \sin\theta} \der_\theta
              (A \sin\theta \der_\theta g)=0, \nn \\
&& A\equiv m^2\sec^2 \dfrac{\theta}{2}+n^2\csc^2\dfrac{\theta}{2}.
\eea
 The equation admits
separation of variables
\bea
&& g(\rho, \theta)=R(\rho) Y(\theta).
\eea
A solution for the radial function is given by
\bea
&& R(\rho)=C_1 \sin[w^2 \log (\mu \rho)]+C_2 \cos[w^2 \log (\mu \rho)],\nn \\
&&
\eea
where $w^2$ is the separation constant and $\mu$ is a free mass dimensional parameter.
The equation for the angular function $Y(\theta)$
\bea
&& \dfrac{4}{A\sin\theta} \der_\theta
              (A\sin\theta \der_\theta Y(\theta))-w^2 Y(\theta)=0.
\eea
can be reduced to the Heun type equation.
Notice, the angular part of our solution, $Y(\theta)$,
is the same as in \cite{ferr2004}, however,
the radial function $R(\rho)$ is different.
An essential difference is that our solution has a dimensional mass scale
parameter $\mu$ which is completely arbitrary contrary to
the case of the solution with a fixed radius $r_0$
in the $CP^1$ model on the space with topology
$S^3 \times R^1$.

\section{Knot ansatz for solutions in QCD in Euclidean space-time}

One method of constructing solutions is based on deformation
of vacuum configurations using an appropriate ansatz with suitable trial functions.
Since the classical vacuum can be described in terms
of the $CP^1$ vector field $\hat n$ which possesses non-trivial topological properties,
an interesting possibility arises: to obtain new solutions starting with
known exact Hopfion solutions in $CP^1$ integrable model. Another advantage
of using the field $\hat n$ with non-trivial topology in studying vacuum
excitations appears when one considers $SU(3)$ QCD which has more rich
topological structure.
In this section we apply $(m,n)$-family of known static solutions
$\hat n$ \cite{ferr2004} to construction of new classical solutions
in Euclidean $SU(2)$ QCD.

We define a following ansatz with three radial trial functions
$f_i(\rho)$
\bea
\vec A_\mu=- f_1(\rho)\tC_\mu \hat n-f_2(\rho) \hat n \times \der_\mu\hat n
+f_3(\rho) \der_\mu \hat n,
 \label{instans}
\eea
where $\hat n$ describes Hopf mapping with Hopf charge $Q_H=mn$
and it is defined by the complex function
\bea
u=\exp[i(m \phi +n \psi)] g(\theta; m,n).
\eea
We consider a simple case when the function $g(\theta; m,n)$
is defined for equalled winding numbers $m=n$ \cite{AFZ}
\bea
g(\theta; m,n)=\cot\frac{\theta}{2}.
\eea
Substituting the ansatz into the full equations of motion of pure QCD
\bea
\vec D_\mu \vec F^{\mu\nu}=0
\eea
one can obtain solutions with finite and infinite energy.
We select the most interesting solutions:

1. 't Hooft instanton with Chern-Simons number $N_{CS}=1$:
with the constraints $m=n=1,~f_3(\rho)=0,~ f_1(\rho)=f_2(\rho)$
all equations of motion reduce to one non-trivial ordinary differential equation:
\bea
&&r^2 f''_1+r f_1'-4f_1(f_1-1)(2 f_1-1).
\eea
The equation has a simple solution
\bea
&&f_1=\dfrac{a^2}{a^2+\rho^2}
\eea
which leads to gauge equivalent representation of known 't Hooft instanton.
One can check that the Chern-Simons number for the corresponding
gauge potential equals exactly to the Hopf number.
Notice, that there are several gauge equivalent representations for the
known 't Hooft instanton which can be obtained within the ansatz
(\ref{instans}).

2. Infinite energy solution with the vector field $\hat n$
with unit Hopf charge: the solution is obtained by using constant valued trial
functions $f_{1,2,3}$. Direct solving all equations of motion leads
to the following solution
%
\bea
&&f_1=\dfrac{1}{2},~~~~~
 f_2=1\pm\dfrac{\sqrt 2}{4},~~~~~ f_3=\pm\dfrac{\sqrt 2}{4}.
\eea
The corresponding energy density (density of the Euclidean action)
takes the form
\bea
&&{\cal E}(\rho)=\dfrac{3 \sin \theta}{8 \rho}
\eea
which coincides exactly with the expression for the energy density of the knot
solution with unit Hopf charge on three-dimensional
sphere with a finite radius $\rho$. In this case it becomes
evident that an exact solution of the $CP^1$ model on three dimensional sphere
turns into the singular solution of the Yang-Mills theory.

3. Singular solutions with the vector field $\hat n$ with
the Hopf charge $Q_H=m^2$ ($m=n$): the ansatz includes two
constant valued trial functions $f_2=1,~ f_3=0$.
In that case all equations of motion reduce to one ordinary differential equation
\bea
m (r^2 f_1''+r f_1'-4(f_1-1))=0
\eea
which has the following solution with an arbitrary parameter
$a$ of length dimension
\bea
f_1=1+c_1 \cosh(2 \log (\frac{r}{a}))+c_2 \sinh (2 \log (\frac{r}{a})).
\eea
The energy density of that solution reads
\bea
{\cal E}(\rho)=m^2 \dfrac{r^3 \sin\theta}{a^4}\Big ((c_1+c_2)^2+\dfrac{a^8 (c_1-c_2)^2}{r^8}\Big ).
\eea
In special cases $c_1=\pm c_2$ the solution takes a simple polynomial form.

We have considered a simple ansatz with spherically symmetric trial functions.
It would be interesting to apply the ansatz with axially symmetric
trial functions to study new finite energy solutions, especially
to the search of essentially $SU(3)$ instanton like solutions.

\section{Singular solutions in QCD in Minkowski space-time}

It is known that Wu-Yang monopole represents a solution to classical
equations of motion in $SU(2)$ Yang-Mills theory.
Within the formalism of the restricted QCD the Wu-Yang monopole
is given by the unit color vector directed along the radius
\bea
\hat n^i=\dfrac{x^i}{r} ~~~~~(i=1,2,3).
\eea
The solution is singular, and it satisfies the equations of motion
$\der^\mu H_{\mu\nu}=0$ everywhere except the origin.
The unit monopole charge of Wu-Yang monopole
is provided by the nontrivial homotopy group $\pi_2(SU(2)/U(1))$.

Let us consider a vector $\hat n$ with a natural knot topology
induced by the definition of the three-dimensional sphere
obtained by compactification of the space $R^3$ to three-sphere $S^3$
by identifying all points at infinity.
To do this we apply stereographic projection of $S^3$ to $R^3$
\bea
x_1&=& \dfrac{2 a^2}{a^2+r^2} X, \nn \\
x_2&=& \dfrac{2 a^2}{a^2+r^2} Y, \nn \\
x_3&=& \dfrac{2 a^2}{a^2+r^2} Z, \nn \\
r^2 &\equiv& X^2+Y^2+Z^2,
\eea
where $X,Y,Z$ are Cartezian coordinate in $R^3$.
The three-dimensional sphere $S^3$ is defined by the equation
\bea
&& x_1^2+x_2^2+x_3^2+x_4^2=a^2,
\eea
where we keep the radius $a$ of the sphere as a free parameter which
will be useful in further.
One can pass from the Cartesian coordinate system $X,Y,Z$  to
the standard spherical coordinates $r,\theta, \phi$.
The complex coordinates $z_1, z_2$ can be written as follows
\bea
&& z_1=x_1+i x_2=\dfrac{2a^2r\sin \theta e^{-i \phi}}{a^2+r^2}, \nn \\
&& z_2=x_3+i x_4=\dfrac{a(2 a r \cos \theta-i (a^2-r^2))}{a^2+r^2}. \label{z1z2}
\eea
It is convenient to define the Hopf mapping by
\bea
 u&=& \dfrac{z_2}{z_1}=
\dfrac{2ar \cos \theta-i(a^2-r^2)}{2ar \sin \theta} e^{i\phi}.  \label{nicole1}
\eea
One can write down an explicit expression for
the color vector $\hat n$ as follows
\bea
\hat n &=&\dfrac{1}{(a^2+r^2)^2}\cdot \nn \\
         &&\left (\ba{c}
  4ar(2 ar \cos \phi \cos \theta+(a^2-r^2) \sin \phi)\sin \theta  \\
 4ar((-a^2+r^2)\cos\phi+2ar\sin\phi\cos\theta)\sin\theta\\
(a^2-r^2)^2+4a^2r^2 \cos(2\theta) \\
            \ea
           \right ). \nn \\
\eea
Notice, that magnetic field components have simple
expressions and correspond to magnetic helical vortex configuration
\bea
H_{r\theta}&=&-\dfrac{32a^3r^2 \sin \theta}{(a^2+r^2)^3}, \nn \\
H_{\theta\phi}&=&-\dfrac{8 a^2r^2 \sin (2\theta)}{(a^2+r^2)^2}, \nn \\
H_{r\phi}&=&-\dfrac{8a^2 r(a^2-r^2)(1-\cos(2\theta))}{(a^2+r^2)^3}.
\eea
Surprisingly, even though the magnetic field is axially symmetric,
the energy density ${\cal E}$ is spherically symmetric \cite{nicole},
and the knot configuration has a finite energy
\bea
E=\int d^3V {\cal E}=\int r^2 \sin \theta dr d\theta d\phi
\dfrac{256 a^4}{(a^2+r^2)^4}=\dfrac{32 \pi^2}{a}. \nn \\
\eea
The solution was first found as an exact solution
in the Nicole model with a Lagrangian \cite{nicole}
\bea
{\cal L} \simeq -((\der_\mu \hat n)^2)^{3/2}, \label{nicolemodel}
\eea
and later as a special solution with winding numbers $m=n=1$
in the integrable Aratyn-Ferreira-Zimerman (AFZ) model
defined by the Lagrangian \cite{AFZ}
\bea
{\cal L}_{AFZ}=-\dfrac{1}{4} (H_{\mu\nu}^2)^{3/4}. \label{AFZmodel}
\eea

We introduce the following axially symmetric ansatz for
the restricted gauge potential
\bea
\vec A_0^a&=&0, \nn \\
\vec A_r^a&=& P(r,\theta) \hat n^a +\vec C_r^a, \nn \\
\vec A_\theta^a&=& Q(r,\theta) \hat n^a +\vec C_\theta^a, \nn \\
\vec A_\phi^a&=& R(r,\theta) \hat n^a +\vec C_\theta^a,
\eea
where $P,Q,R$ are trial functions. Due to the dual $U(1)$
symmetry one can impose a gauge condition on the trial functions.
It is suitable to choose a condition $P(r,\theta)=0$.
The ansatz implies the following expression for the scalar
magnetic field ($m,n=r,\theta,\phi)$
\bea
&&H_{mn}= H_{mn}^a \hat n^a,  \nn \\
&& H \equiv dx^m \wedge dx^n H_{mn}=\nn \\
&& dr\wedge d\theta \Big (\dfrac{32a^3 r^2 \sin \theta}{(a^2+r^2)^3}+Q_r \Big ) + \nn \\
&& dr\wedge d\phi \Big (\dfrac{16 a^2 r (r^2-a^2)\sin^2 \theta}{(a^2+r^2)^3}+R_r \Big ) + \nn \\
&& d\theta\wedge d\phi \Big (-\dfrac{8 a^2 r^2 \sin (2\theta)}{(a^2+r^2)^3}+R_\theta \Big ).
\eea
After substituting the ansatz into the equations of motion
$\vec D^\mu \vec F_{\mu\nu}=0$
one can find that all nine equations reduce
to three linear partial differential equations
\bea
&&(a^2+r^2)^3(\sin \theta  Q_{r\theta}+\cos \theta Q_r)+32 a^3 r^2 \sin(2 \theta)=0, \nn \\
&&(a^2+r^2)^4 Q_{rr}+64 a^3 r (a^2-2 r^2) \sin \theta =0, \nn \\
&&(a^2+r^2)^4 \big (\sin \theta (r^2 R_{rr}+ R_{\theta\theta})-\cos \theta  R_\theta \big) \nn \\
&& ~~~~~-32 a^2 r^4 (-5 a^2+r^2) \sin^3 \theta =0.
 \eea
The fact that we have finally linear differential equations
is caused by Abelian structure of the chosen ansatz.
The solution to the first two equations is given as follows
\bea
Q&=&c_1 +c_2 \dfrac{r}{\sin \theta}-\dfrac{4ar(a^2-r^2)}{(a^2+r^2)^2} \sin\theta \nn \\
 &&-4 \arctan (\dfrac{r}{a}) \sin \theta.
 \eea
A solution to the equation for the function $R$
can be found as a sum of a general solution $R_0 (r,\theta)$
to the homogeneous part of the equation
and a special solution $R_1(r,\theta)$ to the inhomogeneous equation
\bea
R(r,\theta)&=& R_0 (r,\theta) +R_1 (r,\theta), \nn \\
R_1(r,\theta) &=&\big (\dfrac{c_{01}}{r}+c_{02} r^2+\dfrac{8 a^2 r^2}{(a^2+r^2)^2 }\big )
\sin^2 \theta.
\eea
The homogeneous equation for $R_0(r,\theta)$ is separable
and the general solution is given by
\bea
&&R_0(r,\theta)= \sum_{w} K(r;w) Y(\theta ; w), \nn \\
&&K(r ; w) = a_w r^{(1-\sqrt {1+4 w})/2}+b_w r^{(1+\sqrt {1+4 w})/2}, \nn \\
&&Y(\theta ;w)=c_w~ {}_2F_1 {[-p_+,-p_-,1/2;\cos^2\theta]}+ \nn \\
&& ~~~~~d_w \cos \theta~{}_2F_1[p_-,p_+,3/2;\cos^2\theta], \nn \\
&& p_\pm=(1\pm\sqrt {1+4 w})/4,
\eea
where $w$ is a constant of separation.
The hypergeometric functions $~_2F_1(a,b,c;z)$ in the above equations reduce to
trigonometric polynomials for integer and semi-integer values
of $p_\pm$, i.e., if we impose the following conditions
\bea
&&w=l(l+1), ~~~ l=0,1,2,...~, \nn \\
&&c_l=0,~~ {\rm for}~~ l=2,4,6,...~, \nn \\
&&d_l=0, ~~{\rm for} ~~l=1,3,5,...~.
\eea

In general, the solution represents a class of
singular infinite energy field configurations due to singularities
at the origin, $r=0$, or at space infinity, $r=\infty$.
Let us consider special solutions which have singularities
at the origin and the energy density $E(r)$ falling down at infinity
as $\Big (\dfrac{1}{r}\Big )^{\alpha >1}$ ($l=0$):

{\bf (i)} $c_1=c_2=c_{01}=c_{02}=b_w=0$, $a_w=d_w=1$: the solution represents
the Wu-Yang monopole.

{\bf (ii)} $c_1=c_2=c_{02}=b_w=0$, $a_w=d_w=1$: a singular monopole solution
with two non-zero magnetic components $H_{\theta\phi}, H_{r \phi}$.

{\bf (iii)} if $c_2 \neq 0$: all components of the magnetic field $H_{mn}$
are not vanishing. In this case one has an additional singularity along $OZ$ axis.

Notice, since the obtained exact solutions are singular, they are not of much interest
by themselves, especially in the pure QCD. However, the singular solutions
can be used in constructing finite energy solutions in extended theories.
One well-known example is given by the Skyrmion, which can be treated
as a dressed Wu-Yang monopole solution \cite{choprl01}.
Another interesting direction is to consider a more general configuration
for the vector $\hat n$ in search of possible finite energy monopole like
configurations in QCD and $CP^1$ model. This will be considered in
a separate paper \cite{ZZPcp1}.

\section{Skyrme type models with knot solitons}

The knot soliton given by the $CP^1$ field (\ref{nicole1}) represents an exact solution
with a unit Hopf charge in the integrable models \cite{nicole,AFZ} with Lagrangians
(\ref{nicolemodel},\,\ref{AFZmodel}) containing
fractional degrees of the kinetic terms.
These models are far from realistic
physical theories like the standard and effective low energy QCD.
An interesting question arises, is there a Skyrme type model
of effective low energy QCD which
admits a finite energy exact solution with the knot configuration (\ref{nicole1}).
One interesting model was found in \cite{ferr2009} where a generalized
Faddeev-Skyrme model was suggested as a possible low energy effective theory of QCD.
In particular, it had been found that
the Lagrangian
\bea
&& {\cal L}=-\dfrac{1}{4} H_{\mu\nu}^2+ \xi (\der_\mu \hat n)^4 \label{monLagr}
\eea
admits existence of analytic solutions with zero energy in the critical case $\xi=1$.
It turns out that there is another family of Skyrme type models
with an additional scalar field which admits exact analytic
solutions for non-vanishing Hopf numbers.

Let us consider the following Lagrangian
\bea
{\cal L}(\phi, \hat n)=-\mu^2 (\der_\mu \phi)^2-\dfrac{\beta}{4} \phi^2 (\der_\mu \hat n)^2-
\dfrac{\nu}{32} \dfrac{1}{\phi^2} H_{\mu\nu}^2.   \label{Skymod}
\eea
The term proportional to $\dfrac{1}{\phi^2}(\der_\mu \hat n)^4$ can be added
to the Lagrangian as well.
We will not consider this case since the structure of the exact solution
is not affected in principal by adding such a term.
Up to change of variables the first two terms in (\ref{Skymod}) coincide
with the respective first two terms in the original Skyrme model.
The scalar field $\phi$ can be interpreted as a
fluctuation of the length of the field $\hat n$.

The exact solution in AFZ model (\ref{AFZmodel})
in the case of equalled winding numbers $m=n$
has the following form \cite{AFZ} (in toroidal coordinates $(\eta,~\xi,~\phi)$)
\bea
&&u(\eta, \xi,\phi)=\dfrac{1}{\sinh \eta} \exp[i m (\xi+ \phi)]. \label{AFZsol1}
\eea
The Hopf charge of the solution is
given by $Q_H=m^2$. The solution with Hopf charge
$Q_H=1$ coincides with the knot configuration (\ref{nicole1}) in spherical coordinates.
We have found for the case $m=1, ..., 5$ that
for each topological charge $Q_H$ there is a special
Lagrangian (\ref{Skymod}) with constrained parameters $\mu^2, \beta,\nu$
which admits exact finite energy solutions.
The solution to Euler-Lagrange equations
for the scalar field $\phi$ is the same for various $Q_H$
\bea
\phi=\dfrac{a}{\sqrt {a^2+r^2}}.
\eea
The equation of motion for the field $u$ has the following
solution
\bea
u&=&\dfrac{e^{i m \phi}}{2ar \sin \theta}(2ar \cos \theta-i(a^2-r^2)) \cdot \nn \\
&&\Big ( \dfrac{2ar \cos \theta-i(a^2-r^2)}{\sqrt {(a^2+r^2)^2-4 a^2 r^2 \sin^2 \theta}}\Big )^{m-1}
 \label{AFZmm}
\eea
under the assumption that the initial parameters $\mu, \beta,\nu$ in the Lagrangian satisfy
certain constraints.
The solution (\ref{AFZmm}) coincides with the solution (\ref{AFZsol1}) given in toroidal coordinates
up to symmetry $z_1 \leftrightarrow z_2$ in the definition of the Hopf mapping.

 For the case of a unit Hopf charge, $Q_H=1$,
one has a constraint
\bea
\dfrac{3}{8} \mu^2+\beta-\dfrac{\nu}{a^2}=0.
\eea
In the special case of $\nu=0$ one has a model without
any dimensional parameters while possessing
a static solution which includes a dimensional parameter $a$.
The solution represents a saddle point configuration with spherically symmetric
energy density
\bea
{\cal E}_0= \dfrac{a^2 \mu^2 (r^2-3 a^2)}{(a^2+r^2)^3} {\upsilon},
\eea
where $\upsilon=r^2 \sin \theta$ is the integration measure.
One can verify that the total energy after integration
over $r$ vanishes identically. The solution is rather unphysical since
the Hamiltonian is not positively defined.
Another special solution can be selected by an additional
condition, $\beta=0$. In this case
the energy density becomes positively defined
\bea
{\cal E}_1=\dfrac{8\nu (r^2+3 a^2)}{3 (a^2+r^2)^3}\upsilon.
\eea
This implies a finite total energy
\bea
E_1=\dfrac{4 \pi^2 \nu}{a}.
\eea

For Hopf charges $Q_H >1$ corresponding to
winding numbers $m=2,3,4,5$
the solution (\ref{AFZmm}) satisfies the equations of motion
if only the following constraints are imposed:
\bea
&& \beta=0, ~~~~\mu^2- \dfrac{8 m^2\nu}{3a^2}=0.
\eea
The energy density is
\bea
{\cal E}_m=\dfrac{8 m^2\nu (r^2+3 a^2)}{3 (a^2+r^2)^3}\upsilon.
\eea
The total energy of the solution corresponding to
each model with a Lagrangian specified by parameters
$\beta, \mu,\nu$ is proportional to the Hopf charge
\bea
E_m=\dfrac{4\pi^2\nu}{a} Q_H. \label{entot}
\eea
For the case $m \neq n$
we expect that knot solutions in such models will have a total energy
expressed by the same equation (\ref{entot}).
A detailed study of knot solutions in a generalized Skyrme-Faddeev model
with the length parameter $a$ determined by equations of motion
is presented in \cite{ZZP13}.

\section{Discussion}

In standard approach to description of the classical
vacuum in QCD in terms of pure gauge connection
all topologically non-equivalent vacuums
are classified by Chern-Simons number. Using an explicit
construction of the classical vacuum in terms
of the $CP^1$ vector field $\hat n$ we have demonstrated
that classical vacuum possesses a more rich topological structure.
Namely, all non-trivial topological vacuums are described
by two homotopy groups $\pi_{2,3}(SU(N)/H)$.
In \cite{weylsym12} we have studied the Weyl symmetric structure
of the classical QCD vacuum described by the second homotopy group
which determines the monopole charge. In the present paper we consider the
knot topology of the classical vacuum determined by the third homotopy group.
A natural question arises on possible physical manifestations of such topological
vacuum structure. We have considered excitations over the vacuum with a
non-trivial Hopf charge in search of new analytic classical solutions
in the $CP^1$ model, standard QCD and Skyrme type low energy effective theory
of QCD. Our approach of studying new solutions can be useful in constructing
essentially $SU(3)$ topological solitons like instantons and monopoles.
It is interesting to study possible physical
applications of the proposed Skyrme type model which can be relevant
to the low energy effective theory of QCD like the Faddeev-Skyrme model \cite{FNnature}.

Certainly, since QCD is essentially a quantum theory,
it is of great importance to find physical effects of the non-trivial
topological structure at quantum level where the degeneracy of the classical vacuum
should disappear. One possible effect includes a non-trivial Weyl symmetric structure of
minimums in the two-loop effective potential in the presence
of monopole condensates. It would be interesting as well to study
implications of knot and monopole configurations, in particular,
the effects related with contribution of all possible non-trivial topological field
configurations to physical quantities like Wilson loop functional and effective action.
One of principal unresolved problems is how to construct in pure QCD (without introducing
any additional scalar fields) finite energy monopole field which is the main ingredient part of the
mechanism of confinement based on dual color superconductivity model.
Some of these issues will be considered in the forthcoming paper \cite{ZZPcp1}.


\acknowledgments

Authors are grateful to Prof. Y. M. Cho for numerous discussions.
One of authors (D.G.P.) thanks E. N. Tsoy and R. M. Galimzyanov for
useful discussions. The work is supported by
NSFC (Grants 11035006 and 11175215), CAS (Contract No. 2011T1J31),
and by UzFFR (Grant F2-FA-F116).

\vspace{2mm}

\end{document}